# Design, Tuning, and Blackbox Optimization of Laser Systems

Jack Hirschman[1,2,*], Randy Lemons[2,3], Minyang Wang[4], Peter Kroetz[5] and Sergio Carbajo[2,6,7,*]

Chirped pulse amplification (CPA) and subsequent nonlinear optical (NLO) systems constitute the backbone of myriad advancements in semiconductor and additive manufacturing, communication networks, biology and medicine, defense and national security, and a host of other sectors over the past decades. Accurately and efficiently modeling CPA and NLO-based laser systems is challenging because of the multitude of coupled linear and nonlinear processes and high variability in simulation frameworks. The lack of fully-integrated models severely hampers further advances in tailoring existing or materializing new CPA+NLO systems. Such tools are the key to enabling emerging optimization and inverse design approaches reliant on data-driven machine learning methods. Here, we present a modular start-to-end software model encompassing an array of amplifier designs and nonlinear optics techniques. The simulator renders time- and frequency-resolved electromagnetic fields alongside essential physical characteristics of energy, fluence, and spectral distribution. To demonstrate its robustness and real-world applicability–specifically, reverse engineering, system optimization, and inverse design–we present a case study on the LCLS-II photo-injector laser, representative of a high-power and spectro-temporally non-trivial CPA+NLO system.

## 1 Introduction

The 2018 Nobel Prize in Physics was awarded to Donna Strickland and Gerard Mourou for their work on chirped pulse amplification (CPA) of optical pulses[1]. Since then, CPA systems have driven and rapidly expanded the fields of high-power lasers, ultrafast lasers, and nonlinear optics (NLO) and now touch every realm of our lives with applications in medical and biomedical imaging and treatment[2–8]; precision machining in manufacturing, especially for semiconductors and electronics[9–13]; current and future telecommunications[14–17] and terahertz wireless and quantum communications[18–20]; national defense, security, and remote monitoring[21–26]; as well as basic sciences including fusion sciences[27], novel accelerators[28–30] and energy sciences[31–35], and strong-field physics[36–38]. Nevertheless, the ground-up design of CPA systems is non-trivial, especially when cascaded nonlinear subsystems are involved. Furthermore, optics and photonics are in the midst of a massive transformation as a host of machine learning (ML) techniques merge with the field[39–41], promising to revolutionize current approaches to the design and manufacturing of laser systems across academic, industrial and medical, and defense complexes.

[1]Department of Applied Physics, Stanford, 348 Via Pueblo, Stanford, CA 94305, USA. [2]SLAC National Accelerator Laboratory, 2575 Sand Hill Rd, Menlo Park, CA 94025, USA. [3]Colorado School of Mines, 1500 Illinois St, Golden, CO 80401, USA. [4]Department of Statistics, UCLA, 520 Portola Plaza, Los Angeles, CA 90095, USA. [5]Atomically Resolved Dynamics Division, Max Planck Research Department for Structural Dynamics, University of Hamburg, 20146 Hamburg, Germany. [6]Electrical and Computer Engineering Department, UCLA, 420 Westwood, Los Angeles, CA 90095, USA. [7] Physics and Astronomy Department, UCLA, 475 Portola Plaza, Los Angeles, CA 90095, USA.
[*]E-mail(s): jhirschm@stanford.edu, scarbajo@ucla.edu.

The future of ML-optimized optics and photonics will rely on large amounts of data to train these networks and search algorithms. A comprehensive start-to-end (S2E) software model for these laser systems would be the ideal solution for quickly generating the large quantities of data required. Fortunately, there is a rich history in modeling CPA+NLO systems with Frantz and Nodvik[42] laying the groundwork for modern theory and simulation in laser pulse propagation in amplifiers[43,44]. Early frameworks for nonlinear pulse propagation were based on the unidirectional pulse propagation equation[45,46] and the generalized nonlinear Schrodinger equation (NLSE)[47,48]. Modeling these equations has required significant development in numerical techniques such as Runge-Kutta, Fourier split-step, and other spectral methods[49–51], leading to current state-of-the-art software packages[52]. However, these state-of-the-art packages focus on application- or framework-specific systems in isolation, are not modularized for scalability and reproducibility, and do not provide a means for reverse engineering and inverse design, inadvertently ignoring crucial interconnected aspects of cascaded nonlinear processes that hamper potential advances in optics and photonics brought about by exploratory research. We present an efficient, modular and expandable S2E framework that addresses these critical gaps in modeling integrated, sophisticated CPA+NLO systems consisting of a host of spectro-temporal shaping, amplification, and nonlinear conversion stages. This framework is targeted for an emerging renaissance in ML photonics and optical systems.

**2 Model Operation**
To demonstrate our model's robustness, we take a complex CPA+NLO system (Fig. 1), mimicking the drive laser at SLAC National Laboratory's LCLS-II, the world's most powerful X-ray free-electron laser (XFEL). It comprises a mode-locked oscillator (a), a pulse shaper (b), a CPA regenerative amplifier (RA) (c), and NLO upconversion (d). The LCLS-II photo-injector laser system is configured to use dispersion-controlled nonlinear shaping (DCNS)[33], an NLO technique that combines noncollinear sum-frequency generation (SFG) and second-harmonic generation (SHG) to render temporally-shaped flat-top ultraviolet (UV) pulses.

The mode-locked oscillator is simulated as a Gaussian pulse with pulse duration ($\tau_{0,\text{osc}}$) and central wavelength ($\lambda_{0,\text{osc}}$) where the associated Gaussian spectrum is representative of LCLS-II photo-injector's mode-locked oscillator (see Supplementary Information). The pulse shaper is modeled as a transfer function in the frequency domain. The transfer function offers spectral amplitude shaping with control over notch position ($\lambda_{1,ps}$), notch width ($\delta\lambda_{1,ps}$), and depth ($\varkappa$) and spectral phase shaping with control over first-, second- (SOD), third- (TOD), and fourth- (FOD) order dispersion[53].

The regenerative amplifier implements the modified Frantz-Nodvik (mFN) equations for working across spectral bandwidth[44]. mFN updates the spectral fluence for each successive pass through the amplifier for N number of passes. The same procedure is used for both the pumping and amplification processes (see Supplementary Information). The pump fluence is taken as a narrowband peak centered at $\lambda_P$ with pump power $P_P$ where the gain crystal is pumped in a single pass. The input seed fluence is taken from the input field for the amplifier module and adjusted by the pump and signal mode radius, r. The NLO module models DCNS by solving the wave equation with the slowly varying wave approximation and a Fourier split-step method[33,54].

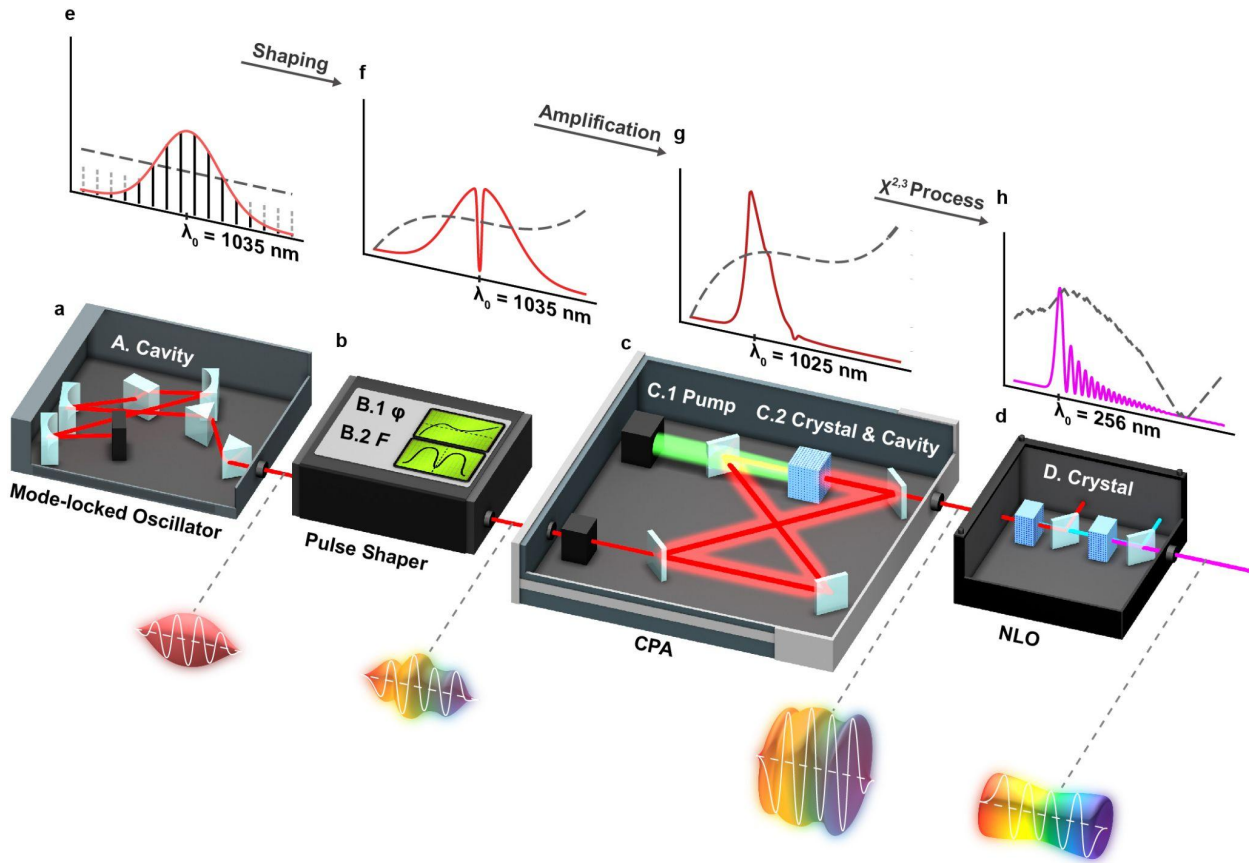

**Fig. 1 | CPA+NLO model overview.** Conceptual diagrams of an oscillator **a**, pulse shaper ($Fe^{i\varphi}$) **b**, regenerative amplifier-based CPA **c**, and cascaded NLO stage **d** accompanied by spectral representations of **e** a mode-locked oscillator, **f** an example of amplitude and phase shaping applied by the pulse shaper, **g** amplified spectrally-shaped pulse, and **h** nonlinear conversion output in the spectrum resulting

from $\chi^{2,3}$ processes. A representation of the time-domain pulse is shown between each block in the simulation.

## 3 Results

As new capabilities of this model, we present three primary use cases: 1) reverse engineering CPA systems; 2) CPA laser optimization; 3) integration of cascaded CPA and NLO stages.

### 3.1 Reverse Engineering

Considered trade secrets, commercial and defense CPA manufacturers do not typically disclose laser design parameters, essentially making their devices blackboxes. While it is not our intention to replicate any commercial products, accurately modeling off-the-shelf devices requires a reasonable understanding of these intrinsic variables since even small variations in some of these parameters–e.g. crystal length and orientation, dopant ion density, and mode radius–significantly alter the amplifier's collective performance. For instance, increasing the dopant ion concentration by 10%, a typical tolerance for crystal manufacturers, may yield a 10% reduction in amplifier spectral intensity bandwidth and a near doubling in output energy in some CPA systems (see Supplementary Information). More generally, the gain medium, concentration of dopant ions, temperature gradients, and mixing of gain crystal axes all significantly affect Stark energy levels, broadening mechanisms, and amplification and emission cross-sections[55–58]. Determining these or even controlling them with high accuracy can be used to mitigate amplification-driven instabilities[58–60], motivating specific material choices or the design of totally new gain media.

In this work, we use a Yb:KGW-based CPA (Ytterbium-doped Potassium Gadolinium Tungstate) medium known for its high absorption and gain, low lasing threshold, broad wavelength coverage, and high thermal conductivity[58,61,62]. To estimate the internal amplifier parameters, we perform a blackbox optimization by searching through the CPA parameter space (see Supplementary Information Table S1) and comparing the simulated output spectral intensity to experimental measurements using normalized root-mean-squared error. We set an additional minimum output energy threshold of 50 uJ and an error-weighting that punishes solutions that saturate too early (see Supplementary Information). Since the exact $Yb^{3+}$ dopant ion concentration, crystal geometry, mode radius, and transmission losses are unknown, we perform a scan over mode radius (0.1 to 0.8 mm), a parameter $\xi$ that combines crystal length and dopant ion density (1E20 to 4E24 $m^{-2}$), and a parameter $\psi$ that represents the percent contribution from each of the three crystal axes and represents the crystal orientation in the amplifier (a-axis 20–100%, b-axis 0–60%, and c-axis 0–60%, see Supplementary Information). Fig. 2a shows the error heat maps for the full scan of mode radius, $\xi$, and for four out of the one-hundred total sets of ψ. The minimum and maximum errors for each heat map are highlighted. Fig. 2b then shows the smallest error from these four heat maps (local minimum) and the largest error from these heat maps (local maximum) as well as the lowest error from the entire parameter scan (global minimum). The associated energy build-up for each is shown in the inset. The local and global minima result in spectral distributions closely resembling the experiment. The global minimum exhibits more gain narrowing and a slight asymmetry in the intensity that closely matches the experiment.

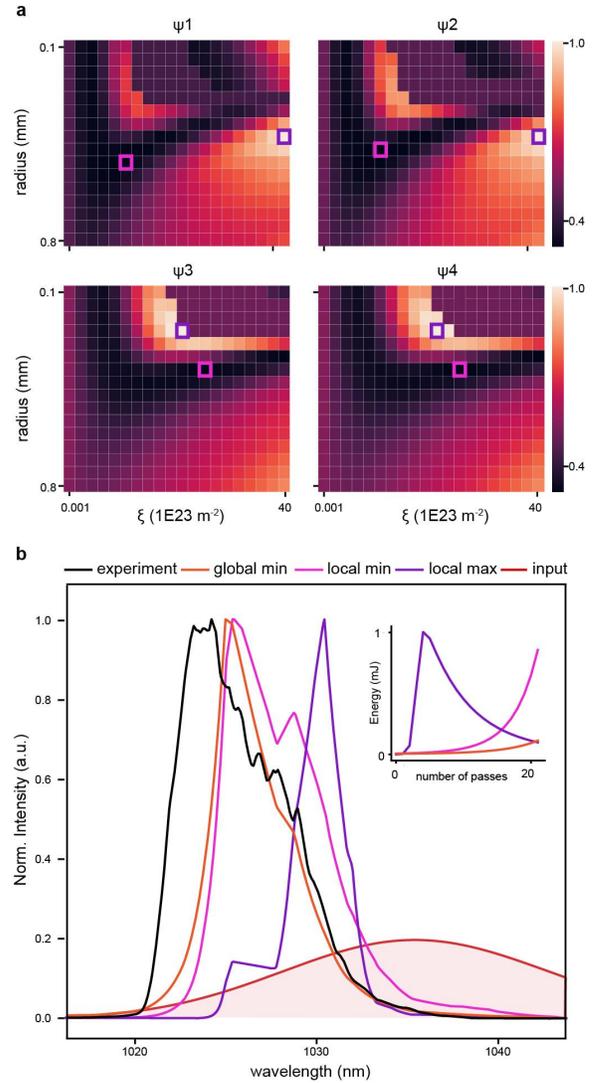

**Fig. 2 | Reverse engineering CPA.** Error heat maps from optimization search in **a** show a subset of searches over beam radius and combined parameter ($\xi$) for crystal orientation parameter (ψ) minimum and maximum error results outlined in magenta and purple, respectively. Post-optimization results in **b** show normalized spectral intensity for input pulse (red), experiment CPA output (black), local minimum and maximum error outputs from **a**, and the global minimum error output from the entire parameter search as well as the corresponding energy build up (inset).

### 3.2 CPA Optimization

Controlling and tuning CPA output spectral bandwidth (BW), central wavelength,

energy, and temporal pulse shape are essential for optimizing ground-up design of amplifiers and for tuning CPA parameters in situ. In an integrated system, these characteristics also affect downstream components. In this use case scenario, the S2E serves as an ideal playground for finding the optimal regimes of operation for the system in question.

It is well-known that pump power and number of passes significantly alter RA-based CPAs but capturing nuanced quantitative trade-offs can be challenging. For instance, taking the Yb:KGW CPA system parameters matched in Section 3.1 via reverse engineering, we showcase control over bandwidth and pulse energy by tuning the number of passes and pump power applied to an initial 50 nJ seed pulse. 10 amplification passes yield a ~10 nm BW and a peak fluence of about 200 $\mu J/m^2$ after amplification. Doubling the number of passes yields a 30% reduction in BW and a 10-fold increase in peak fluence. Bringing the number of passes to 60 then reduces the BW by another 60%, bringing the total output energy up to 3.5 mJ. Doubling the pump power in the CPA with 60 passes leads to a slight decrease in BW and decreases total energy to ~2.9 mJ (Fig. 3). This energy reduction with an increase in pump power comes naturally from saturation of the gain medium and the intra-cavity RA losses.

The gain narrowing is more pronounced at higher pump power yielding higher energy pulses with less number of passes (also see Supplementary Information Fig. S2). Yb-based RAs for example often have around 15-40 round trip passes with tunable ranges of repetition rate exceeding 1 MHz, where the pump power can be used to tune and optimize energy while avoiding side effects from many passes, such as limiting repetition rate or accumulating additional nonlinear phase. As seen in Fig. 3e and 3f at low pump power (120 W) saturation is reached around the $60^{th}$ pass. At high pump power (240 W), saturation is reached at the $50^{th}$ pass and amplification efficiency begins to decay thereafter. These are well-understood phenomena and in their own right show how the model framework can generate data with these simulations to inform ML-based search algorithms for CPA system parameter optimization. These results demonstrate a few of the knobs available for tuning the amplifier, especially for in-situ optimization but become even more important when integrated with the rest of the laser system.

### 3.3 CPA and NLO Integration

The S2E model's full capabilities are shown when combined with cascaded nonlinear processes. Incorporating CPA simulation reveals important, often overlooked subtleties in simulating cascaded processes, in particular the amplified spectrum's effect on downstream pulse shape. More importantly, connecting pre-amplification pulse shaping with post-amplification upconverion where the shaping does not map linearly, opens the door to ML-based studies reliant on large amounts of data.

Our example system model from LCLS-II contains a pulse shaper, amplifier, and DCNS upconversion with SFG and SHG. Combining a programmable pulse shaper with a CPA system provides added flexibility in temporal and spectral shaping. The phase of the electric field can be used to change the temporal envelope of a pulse, and current technology can access the first four orders of phase with both programmable elements like spatial light modulators[63,64], acousto-optic devices[64,65], and emerging structured photonics[66–68] as well as static methods including prism-based shaping, grating-based shaping, and sculpting of pulses in fiber[33,69–73]. Altering the spectral envelope also affects the temporal pulse envelope, which can be exploited for amplitude pulse shaping and modulation but can also exhibit unintended consequences like spectral narrowing, spectral distortions, and wavelength shifts. Furthermore, applications such as pump lasers for OPCPA systems[74,75] and photo-injector lasers[76] require higher harmonics with specific or tunable pulse shapes. These can be produced via NLO systems proceeding shaping and amplification[77].

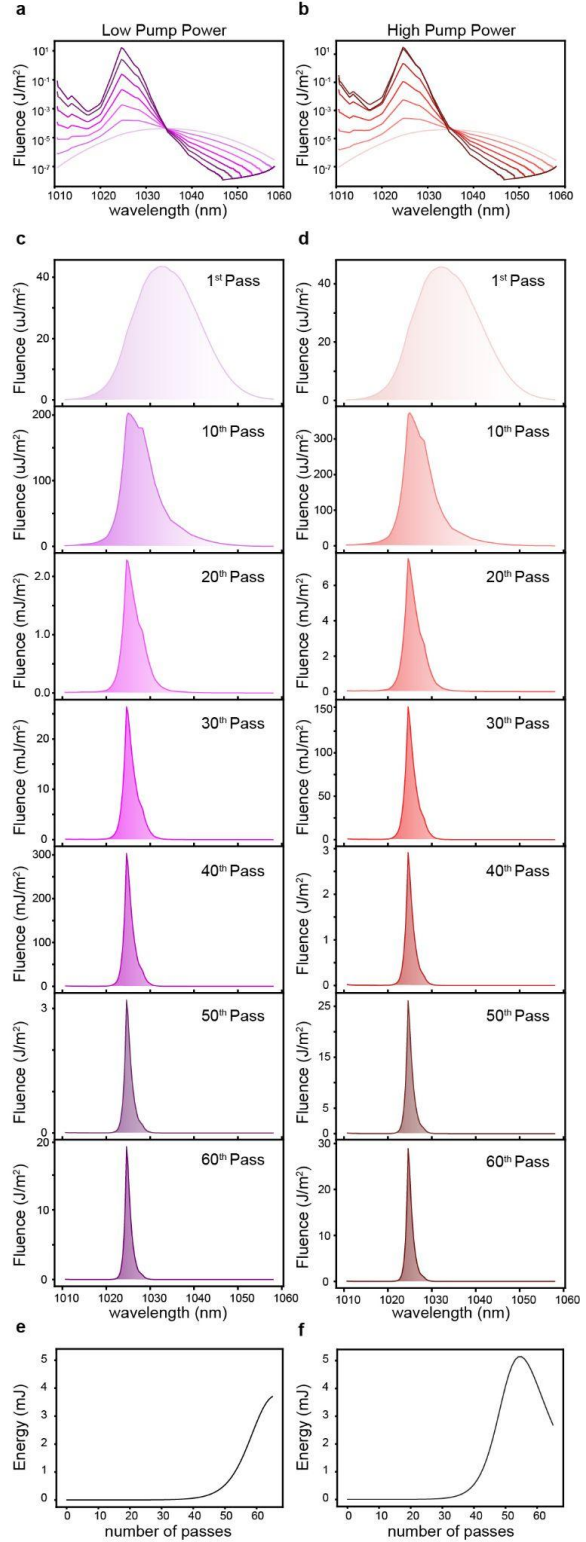

**Fig. 3 | In situ CPA optimization. a** and **b** show the amplified pulse at each pass in the RA overlayed in log scale and **c** and **d** show these pulses separated in linear scale for a pump power of 120 W and 240 W,

left and right respectively. **c** and **d** show the corresponding energy build-up over the passes in the amplifier.

While some of these applications target flat-top pulses post-upconversion, other applications strive for particular time-domain pulse shapes pre-upconversion, such as wakefield accelerators where a triangular-shaped laser pulse can drive a highly sought-after electron-bunch charge distribution[74]. Fig. 4a exhibits this pre-amplification triangle pulse (in light red) and 4e shows the corresponding frequency domain phase required for achieving it.

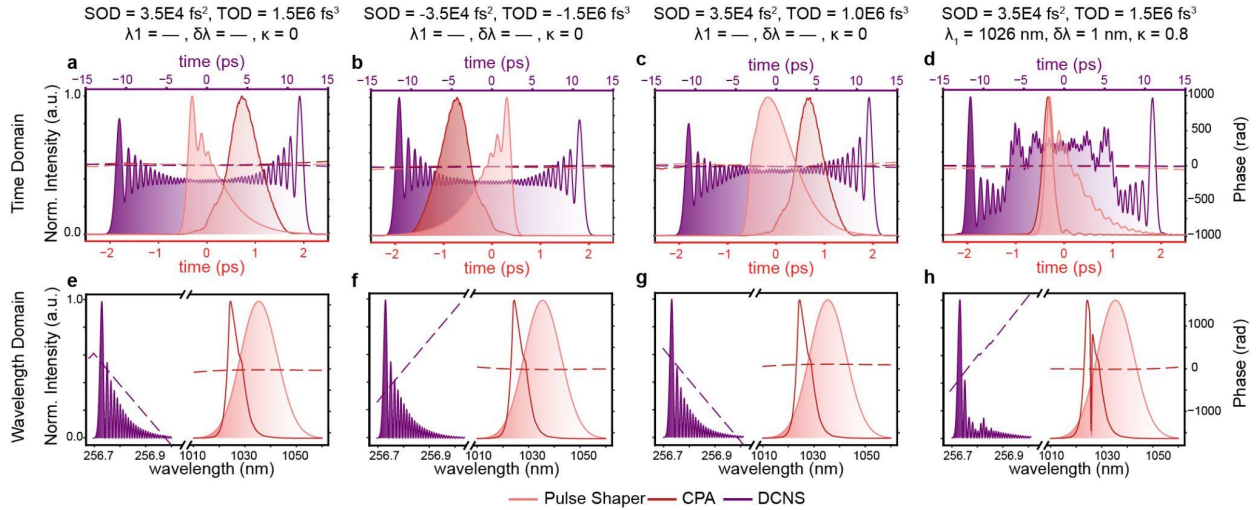

**Fig. 4 | Pulse shaper, amplifier, and upconverted pulses.** Time domain (**a**–**d**) and frequency domain (**e**–**h**) intensity (solid) and phase (dashed) for pulse shaper (light red), CPA (dark red), and DCNS (purple) simulation outputs for various combinations of SOD, TOD, and spectral notches (each column). (**a**–**d**) have a separate time axis for the DCNS signal (top) and pulse shaper and CPA (bottom).

After amplification (in dark red), the time-domain intensity changes drastically from the desired triangle pulse at the output of the pulse shaper, showcasing how full simulation including amplification is necessary in order to understand pulse shape-driven dynamics. The upconverted output pulse from DCNS (purple) remains relatively flat-top but with an asymmetry in the lobe caused by the added phase from the pulse shaper. Shown in Fig. 4b and 4f, flipping the signs of both SOD and TOD reverses the slope of the pulse shaper temporal output, mirrors the amplifier temporal output, and switches the asymmetry's side for the upconverted temporal pulse. Fig. 4c and 4g show how sensitive the dynamics are. With a 33% reduction in TOD compared to Fig. 4a, the triangular pulse at the pulse shaper output begins to smooth and reduce its peak sharpness. The amplified signal change is less noticeable since the modified spectrum due to amplification is a more dominant effect. Similarly, the upconverted pulse shape remains relatively unchanged. Finally, in Fig. 4d and 4f, we introduce a spectral notch and use the same amount of added phase as Fig. 4a and 4e. This notch removes some frequencies resulting in additional oscillations in the pulse shaper and amplifier temporal envelopes. The upconverted temporal profile then exhibits a significantly raised flat-top profile of 10 ps in duration, which could be useful in the LCLS-II example case for enhancing X-ray production[33]. More broadly, this section showcases the ability to combine multiple distinct systems together to examine exploratory pulse shaping effects for applications in light by design or remediation of nonlinear pulse distortion[68]. Given its modular design, this framework facilitates swapping in-and-out various CPA and NLO system models, tuning and optimizing those models for the targeted application, and, ultimately, running full S2E simulations to develop a more complete understanding of the system.

## 4 Conclusions

The advent of the CPA has a track record of dramatic impact on society with high-power laser systems affecting areas as disparate as medical procedures and diagnoses to national defense and security. In order to continue driving progress in the field and to prepare for an age focused around applying data-driven ML to CPA and NLO systems, accurate and efficient fully integrated S2E modeling is crucial. Our S2E framework addresses this critical simulation gap by providing an accurate, modular, and expandable simulator for use cases spanning from reverse engineering and blackbox optimization of hardware components and high-power laser and nonlinear systems, to optimization and tuning of current system designs, and the full-scale simulation of optical systems for data generation. This framework is poised to transform traditional approaches in laser science and engineering into translational discoveries and technologies impacting the design of new photonics and optical devices, impacting light-driven applications in chemical, biological, and medical applications, and accelerating energy sciences in fusion energy sources.